\documentclass{iau} 
\usepackage{graphicx}
\usepackage{natbib}
\usepackage[labelsep=quad,indention=10pt]{subfig}
\title[JD 11.~~Pre-solar grains and AGB stars] 
{A systematic survey of grain growth in \\ discs around post-AGB binaries with PACS and SPIRE photometry}

\author[K. Dsilva, H. Van Winckel \& J. Kluska]   
{K. Dsilva$^1$,
 H. Van Winckel$^1$ \and J. Kluska$^1$}

\affiliation{$^1$Instituut voor Sterrenkunde, KU Leuven, \\ Celestijnenlaan 200D,
3001 Leuven, Belgium \\ email: {\tt karansingh.dsilva@student.kuleuven.be} \\
}

\pubyear{2018}
\volume{343}  
\jname{Why Galaxies Care About AGB Stars}
\editors{F. Kerschbaum, M. Groenewegen, \& H. Olofsson}
\begin{document}

\maketitle

\begin{abstract}
Post-AGB stars are the final stage of evolution of low-intermediate mass stars (M $<$ 8M$_\odot$). Those in binary systems have stable circumbinary discs. Using data from Herschel (PACS/SPIRE), we extend the SEDs of 50 galactic post-AGB binary systems to sub-millimetre wavelengths and use the slope of the SED as a diagnostic tool to probe the presence of large grains. Using a Monte Carlo radiative transfer code (MCMax), we create a large grid of models to quantify the observed spectral indices, and use the presence of large grains in the disc as a proxy for evolution. 
\end{abstract}

\firstsection 
\section{Introduction}
The presence of a binary companion drastically affects the evolution of an AGB star. Mass loss, mass transfer and tidal interaction play strongly impact the outcome of this interaction \citep{pols2004}. Through long-term radial velocity monitoring, it is now well established that post-AGB binary systems are often surrounded by long-lived, stable circumbinary discs \citep{vanwinckel2009,vanwinckel2017}. Interaction between the circumbinary disc and the central system might impact the evolution of the system, and hence it is important to understand the evolution of the disc. The formation and lifetime of the discs are unknown. Here, we present a systematic survey of 50 galactic post-AGB binaries and with the help of radiative transfer models, attempt to use the grain-sizes in the disc as a proxy for evolution.

\section{Methods}
The first step was to obtain accurate photometric fluxes from the PACS \citep{2010pacs} and SPIRE \citep{2010spire} instruments on board the Herschel space telescope. This was done using the recommended software HIPE \citep{2010hipe}.\\
Once the fluxes were obtained, they were added to the respective SEDs of the sources. In order to systematically study the sample, a comparison of the PACS and SPIRE slopes (called the spectral index) were made, called \textit{n$_{160}$} and \textit{n$_{500}$} respectively. Sources with missing fluxes or upper limits were excluded, and we proceeded with 37 targets. \\
Upon comparing \textit{n$_{500}$} and \textit{n$_{160}$}, the relative change $\delta n$ in slope was measured and the sample was classified into three types:
\begin{itemize}
    \item \textbf{Type 1}: Slopes with a knee like feature with $\delta n > 10\%$ (13/37 sources).
    \item \textbf{Type 2}: No change in slope with $\delta n < 10\%$ (12/37 sources).
    \item \textbf{Type 3}: An infrared excess with $\delta n > 10\%$ (12/37 sources).
\end{itemize}

\begin{figure}
    \centering
    \includegraphics[scale=0.4]{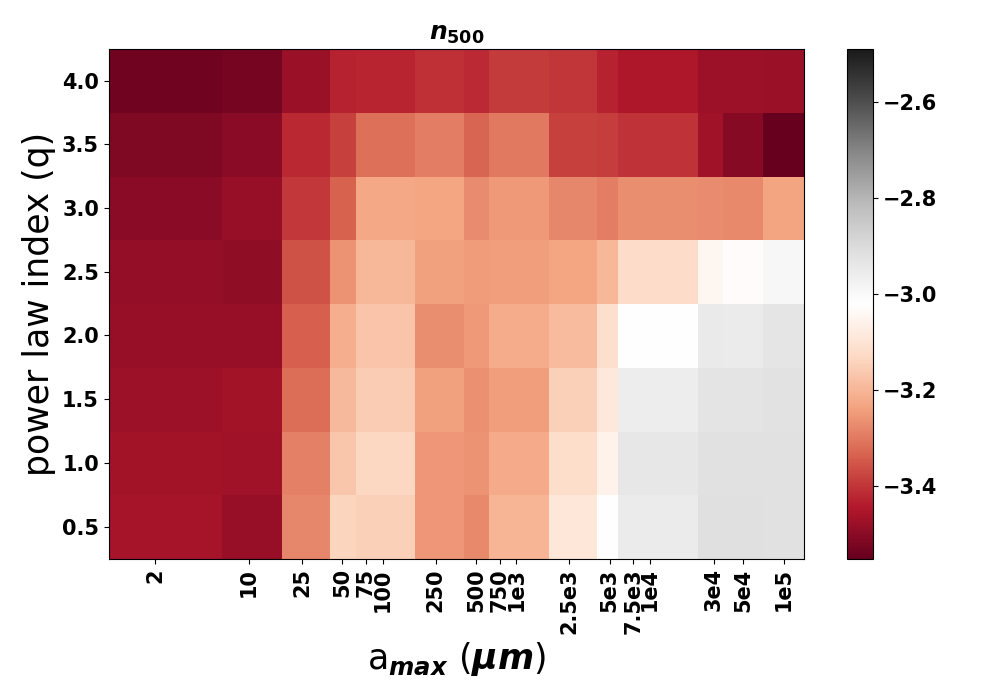}
    \caption{The synthetic spectral indices (n$_{500}$)}
    \label{fig:n500}
\end{figure}

After a parameter study using radiative transfer code MCMax \citep{min2009}, the maximum grain-size (a$_{max}$) and the grain-size distribution power-law index (q) were varied across a fixed input for a second generation disc with parameters adapted from \citet{kluska2018}. The reader is referred to \citet{min2009} for an in-depth explanation of the parameters. The parameter space can be seen in Figure \ref{fig:n500}.

\section{Results and discussion}
Using Figure \ref{fig:n500}, we can compare the observed value of n$_{500}$ to the synthetic spectral index and constrain the appropriate q and a$_{max}$ value for each source.\\  
We observe that the spectral index at 500 is the most sensitive to the presence of large grains and can reliably be used to q and a$_{max}$, which helps in reducing the degeneracies that plague disc modelling.\\
The maximum grain size in all sources is much larger than what is thought to be during the AGB wind phase (a$_{max} >> 0.1 \mu$m). The interpretation made here is that the timescale of grain-growth is relatively short, and hence grain-growth is omnipresent. \\
Many discs show grains of the order of 1mm. Further follow-ups of `pebble'-sized grains in these second-generation discs could help in understanding grain-growth in protoplanetary discs.

\end{document}